\documentclass{article}
\usepackage{ismir_IEEEtran_small_fixedTitle}
\usepackage{amsmath}
\usepackage{amssymb}
\usepackage{cite}
\usepackage{graphicx}
\usepackage{url}
\usepackage{soul}

\title{Robust Joint Alignment of \\ Multiple Versions of a Piece of Music}

\threeauthors
  {Siying Wang} {\\\\}
  {Sebastian Ewert} {Queen Mary University of London, UK \\ {\tt \{siying.wang,s.ewert,s.e.dixon\}@qmul.ac.uk}}
  {Simon Dixon} {\\}

\renewenvironment{thebibliography}[1]{%
   \begin{oldthebibliography}{#1}%
     \setlength{\parskip}{1.4pt}%
     \setlength{\itemsep}{2.8pt}%
   }%
   {\end{oldthebibliography}}

\def\fspace{{\mathcal F}}
\def\R{{\mathbb R}}
\def\gappen{{\mathcal C_G}}
\newcommand{\setmacro}[2]{{$[#1$\,:\,$#2]$}} %

\newcommand{\sebastian}[1]{{}}

\begin{document}
\maketitle

\begin{abstract}
Large music content libraries often comprise multiple versions of a piece of music. To establish a link between different versions, automatic music alignment methods map each position in one version to a corresponding position in another version. Due to the leeway in interpreting a piece, any two versions can differ significantly, for example, in terms of local tempo, articulation, or playing style. For a given pair of versions, these differences can be significant such that even state-of-the-art methods fail to identify a correct alignment. In this paper, we present a novel method that increases the robustness for difficult to align cases. Instead of aligning only pairs of versions as done in previous methods, our method aligns multiple versions in a joint manner. 
This way, the alignment can be computed by comparing each version not only with one but with several versions, which stabilizes the comparison and leads to an increase in alignment robustness.
Using recordings from the Mazurka Project, the alignment error for our proposed method was 14\% lower on average compared to a state-of-the-art method, with significantly less outliers (standard deviation 53\% lower).
\end{abstract}

\section{Introduction}\label{sec:introduction}

Recent years have seen significant efforts to create large, comprehensive music collections.
Music content providers (e.g. Spotify, iTunes, Pandora) rely on their existence, while national libraries and charitable organizations create and curate them in order to provide access to cultural heritage.
For a given piece of music, large collections often contain various related recordings (cover songs, different interpretations), videos (official clip, live concert) and musical scores (in different formats such as MIDI and MusicXML, covering several editions).
To identify and link these different versions, various automatic alignment methods have been proposed in recent years. Such \emph{synchronization methods} have been used to facilitate navigation in large collections \cite{MuellerCKEF10_Sync_ISR}, to implement score following in real-time \cite{DannenbergR06_alignment_ACM, ArztBFFGW14_ScoreFollowingPiano_AES, MontecchioC11_RealTimeSyncViaSequentialMonteCarlo_ICASSP, DuanP11_StateSpaceModelAlignment_ICASSP}, to compare different interpretations of a piece \cite{WidmerDGPT03_Horowitz_AI}, %
to identify cover songs \cite{SerraGHS08_CoverSong_IEEE-TASLP} or to simplify complex audio processing tasks \cite{EwertPMP14_ScoreInformedSourceSep_IEEE-SPM}.

In general, the goal of music synchronization is, given a position in one version of a piece of music, to locate the corresponding position in another version. To compute a synchronization, existing methods align two versions of a piece at a time, even if several relevant versions are available. For example, in \cite{JoderER11_ConditionalRandomFieldSync_TASLP,EwertMG09_HighResAudioSync_ICASSP} a score of a piece is automatically aligned to a corresponding audio recording, while in \cite{DixonW05_MATCH_ISMIR} two acoustic realizations are being synchronized. As shown previously, current methods yield in many cases alignments of high accuracy \cite{JoderER11_ConditionalRandomFieldSync_TASLP,EwertMG09_HighResAudioSync_ICASSP,DixonW05_MATCH_ISMIR}. However, musicians can interpret a piece in diverse ways, which can lead to significant local differences in terms of articulation and note lengths, ornamental notes, %
or the relative loudness of notes (balance). If such differences are substantial, the alignment accuracy of state-of-the-art methods can drop significantly. 

To increase alignment robustness for difficult cases, 
the main idea in this paper is to exploit the fact that multiple versions of a piece are often available and can be aligned in a joint way.
This way, we can exploit the additional information that each version provides about how a certain position in a piece can be realized by a musician.
As a consequence, while two given recordings might be rather different and hard to align, both of them might actually be more similar to a third recording and including such a recording within the alignment process can lead to an increase in overall robustness. To compute our joint synchronization, we modify a multiple sequence alignment method typically employed in biological signal processing and combine it with strategies developed in a musical context based on Multiscale-DTW (FastDTW) and chroma-based onset features for increased computational efficiency and synchronization accuracy.
In the following, we describe technical details of this method in Section~\ref{sec:alignmentMethod}. Then, we report on some of our experiments in Section~\ref{sec:evaluation}. Conclusions and prospects for future work are given in Section \ref{sec:conclusion}.

\section{Alignment Method}\label{sec:alignmentMethod}

Various methods have been proposed to align two given data sequences, including Dynamic Time Warping (DTW) and Hidden Markov Models (HMM) \cite{DannenbergR06_alignment_ACM}, Conditional Random Fields (CRF) \cite{JoderER11_ConditionalRandomFieldSync_TASLP}, and Particle Filter / Monte-Carlo Sampling (MCS) based methods  \cite{MontecchioC11_RealTimeSyncViaSequentialMonteCarlo_ICASSP, DuanP11_StateSpaceModelAlignment_ICASSP}.
With the exception of MCS methods, which are online methods, the remaining three methods operate in an offline fashion and are quite similar from an algorithmic point of view.
We describe our proposed method as an extension to DTW. However, the underlying ideas are applicable in HMM and CRF contexts as well.

\subsection{Baseline Pairwise Alignment}\label{sec:baseline}

\begin{figure}[!t]
 \centering
 \includegraphics[width=\columnwidth]{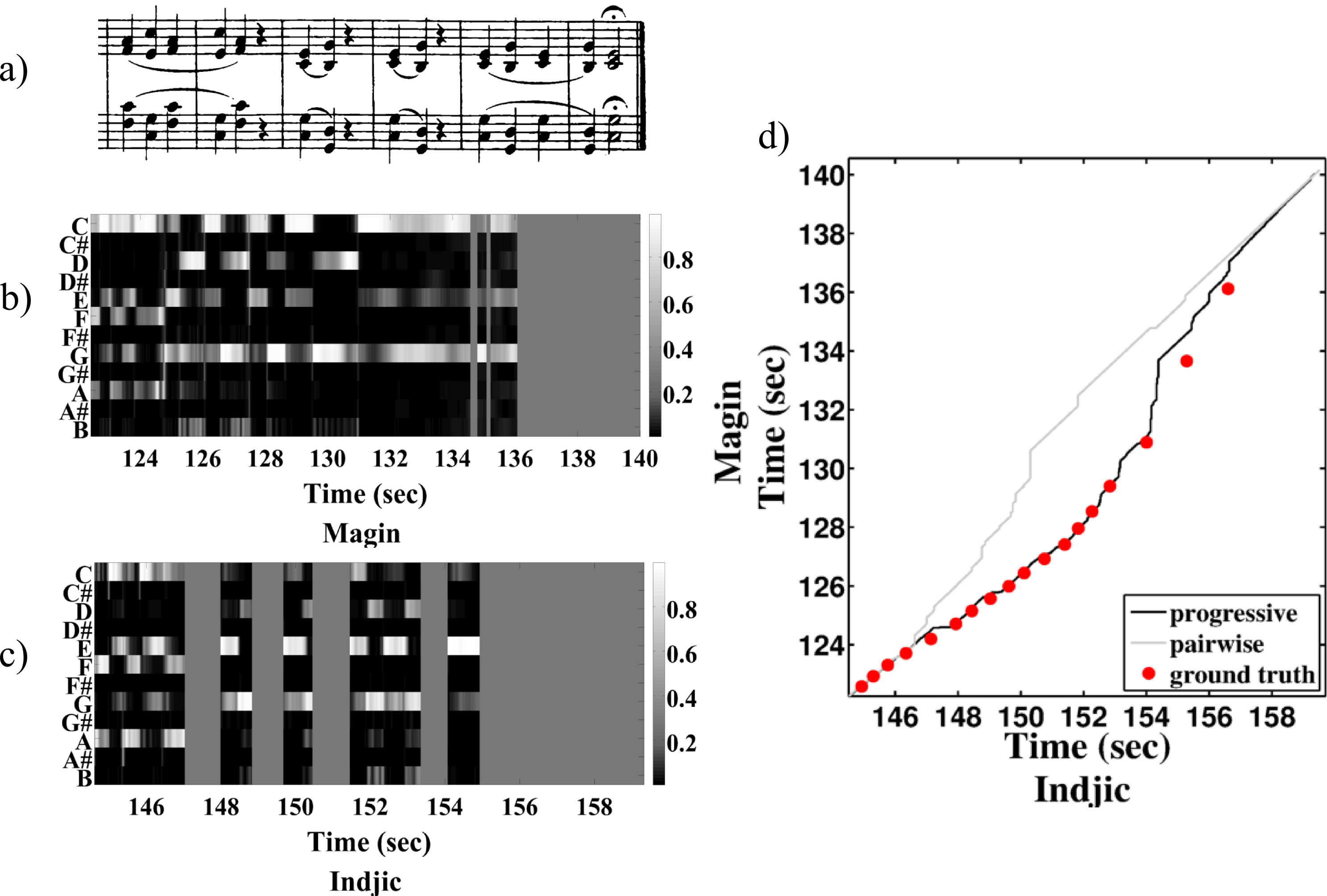}
 \caption{\small Alignment of two interpretations of Chopin Op.~24 No.~2, measures 115-120: \textbf{(a)} Score for the six measures. \textbf{(b)}/\textbf{(c)} Chroma features for an interpretation by Magin and Indjic, respectively; chroma features with uniform energy distribution are the result of silence in the recording. \textbf{(d)} Alignment results for our baseline pairwise (gray) and proposed method (black).}
 \label{fig:Motivation}
\end{figure}

To summarize DTW-based alignment, 
let $X := (x_1,x_2,$ $\ldots,x_N )$ and $Y := (y_1,y_2,\ldots,y_M )$ be two feature sequences with $x_n,y_m \in \fspace$, where $\fspace$ denotes a suitable feature space.
Furthermore, let $c: \fspace \times \fspace \to \R$ denote a local cost measure on $\fspace$.
We define a resulting $(N \times M)$ cost matrix $C$ by $C(n,m) := c(x_n ,y_m)$.
An alignment between $X$ and $Y$ is defined as a sequence $p = (p_1, \ldots, p_L)$ with $p_\ell = (n_\ell,m_\ell) \in$ \setmacro{1}{N}$\times$\setmacro{1}{M}  for $\ell\in$ \setmacro{1}{L} satisfying $1=n_1 \leq n_2 \leq \ldots \leq n_L = N$ and $1 = m_1 \leq m_2 \leq \ldots \leq m_L = M$ (boundary and monotonicity condition), as well as $p_{\ell+1} - p_{\ell} \in \{(1,0),(0,1),(1,1)\}$ (step size condition).
An alignment $p$ having minimal total cost among all possible alignments is called an \emph{optimal alignment}.
To determine such an optimal alignment, one recursively computes
an $(N \times M)$-matrix $D$, where the matrix entry $D(n,m)$ is the total cost of the optimal alignment between $(x_1,\ldots,x_n)$ and $(y_1,\ldots,y_m)$:
\[
D(n,m) := \min
\begin{cases}
D(n - 1,m - 1) + w_1  C(n,m),\\
D(n - 1,m) + w_2  C(n,m),\\
D(n,m - 1) + w_3  C(n,m),
\end{cases}
\]
for $n,m > 1$. Furthermore, $D(n,1) := \sum_{k=1}^n w_2 C(k,1)$ for $n > 1$,
$D(1,m) = \sum_{k=1}^M w_3  C(1,k)$ for $m > 1$, and $D(1,1) := C(1,1)$.
The weights $(w_1 ,w_2 ,w_3) \in \R_+^3$ can be used to adjust the preference over the three step sizes.
By tracking the choice for the minimum starting from $D(N,M)$ back to $D(1,1)$, an optimal alignment can be derived in a straightforward way \cite{DannenbergR06_alignment_ACM}.
In a musical context, $\fspace$ typically denotes the space of normalized chroma features, $c$ is usually a cosine (or Euclidean) distance with weights set to  $(w_1 ,w_2 ,w_3) = (2,1,1)$ to remove a bias for the diagonal direction \cite{DannenbergR06_alignment_ACM,DixonW05_MATCH_ISMIR}.

A main difficulty in aligning music stems from the degree of freedom a musician has in interpreting a score, in particular regarding the local tempo, balance (relative loudness of concurrent notes), articulation and playing style.
If several differences occur together, standard alignment methods sometimes fail to identify the musically correct alignment. In Fig.~\ref{fig:Motivation}(b)/(c), we see chroma features for two interpretations of Chopin Op.~24 No.~2 measures 115-120 (Fig.~\ref{fig:Motivation}(a)) as performed by Magin and by Indjic, respectively. Besides the tempo, we see differences in the interpretation of pauses (the uniform energy distributions in the features correspond to silence), articulation and in the balance (relative loudness of notes). In this case, the differences are significant such that pairwise DTW-based approaches \cite{EwertMG09_HighResAudioSync_ICASSP,DixonW05_MATCH_ISMIR} fail to compute the correct alignment, see upper path in Fig.~\ref{fig:Motivation}(d). The red dots indicate corresponding beat positions in the two versions.

\subsection{Joint Alignment of Multiple Versions}
\label{sec:jointAlignment}

Comparing several versions of a piece, interpretations vary in different ways and to different extents.
If several versions of a piece are available, each version provides an example of how a specific position in a piece can be realized, and this additional information can be used to stabilize the alignment for difficult sections.
A straightforward strategy to compute a joint alignment could be to extend DTW to allow for more than two versions.
For example, to align three versions, one can define an order-3 cost tensor in a straightforward way and apply the same dynamic programming techniques as used in DTW \cite{DurbinEKM99_BiologicalSequenceAnalysis_CUP} (note that a cost matrix for two versions is an order-2 tensor).
However, assuming that each feature sequence to be aligned is roughly of length $N$, the time and memory requirement to align $K$ recordings would be in $O(N^K)$, which prohibits the alignment of more than a very few recordings.

In computational biology, multiple sequence alignment is a well-studied problem. Most popular are so called \emph{profile-based methods} and \emph{progressive alignment methods} \cite{DurbinEKM99_BiologicalSequenceAnalysis_CUP}.
Profile-based methods employ a specific type of HMM, which is trained via Expectation-Maximization (EM) on the set of feature sequences to be aligned. Each state of the resulting \emph{profile-HMM} corresponds to a position in a so called average-sequence: the sequence of means of the observation probabilities of the HMM-states, see \cite{DurbinEKM99_BiologicalSequenceAnalysis_CUP} for details. A multiple synchronization is then computed by aligning each sequence to the average-sequence via the Viterbi algorithm. This procedure has been attempted in a musical context with limited success \cite{Robertson13_CPM_MasterThesis}. We believe this is due to, using EM training, whereby aligned features are essentially averaged (with Gaussian observation probabilites), which results in a loss of information and can lead to a loss of alignment accuracy.

\begin{figure}[!t]
 \centering
 \includegraphics[width=\columnwidth]{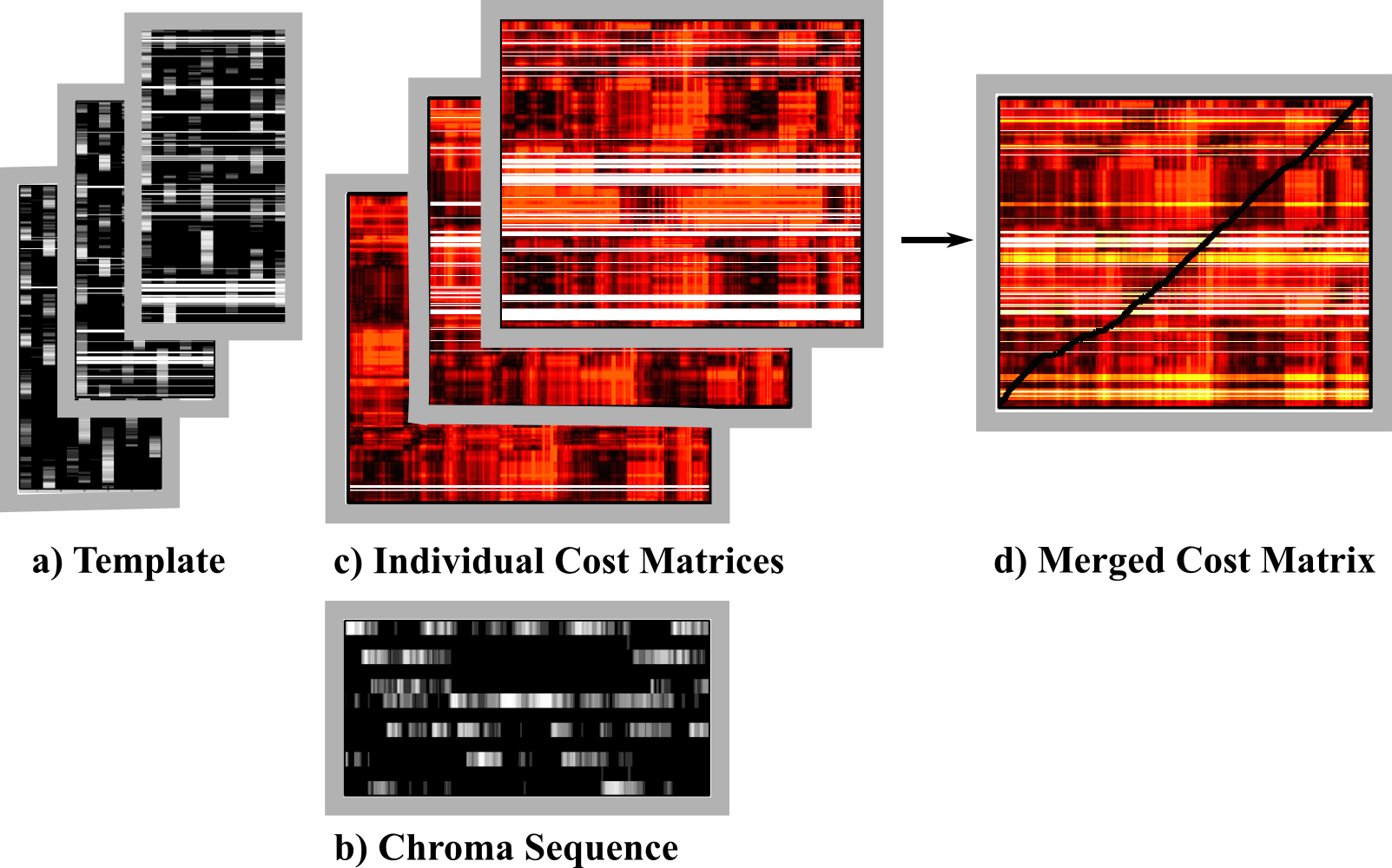}
 \caption{\small Progressive alignment: Three aligned chroma sequences contained in the template \textbf{(a)} are compared to the chroma sequence \textbf{(b)}. The resulting individual cost matrices \textbf{(c)} are merged into one \textbf{(d)}, which is used to compute the alignment. The white lines in (a) and (c) indicate the positions of gap symbols.}
 \label{fig:costMatrix}
\end{figure} 

Using progressive alignment such averaging is not necessary. The underlying idea is to successively build a data structure referred to as a \emph{template}, which provides efficient access to several aligned feature sequences, see Fig.~\ref{fig:costMatrix}(a). By comparing a given feature sequence (Fig.~\ref{fig:costMatrix}(b)) to the sequences contained in the template, the alignment can be computed not only using one cost matrix (as in pairwise alignment) but several matrices in parallel - one for each sequence in the template (Fig.~\ref{fig:costMatrix}(c)).
By suitably combining the information provided by each individual cost matrix, the influence of strong local differences on the alignment, that often only occur between specific pairs of versions, can be attenuated. As shown in Section~\ref{sec:evaluation}, this can lead to a significant boost in alignment robustness.

To describe this procedure in more detail, we assume that we have $K$ different versions of a piece and that their feature sequences are denoted by $X^k = (x^k_1,\ldots,x^k_{N_k})$ for $k\in$\setmacro{1}{K}.
In each step of the progressive alignment, the template $Z$ contains several of these feature sequences that have been stretched to have the same length. Initially, $Z$ only consists of $X^1$. The remaining feature sequences are then successively aligned to $Z$, and after each alignment $Z$ is updated by adding one more sequence. To this end, let $\widetilde{Z} = (\widetilde{z}_1,\ldots,\widetilde{z}_{\widetilde{L}})$ denote the current template which contains $k-1$ sequences of length $\widetilde{L}$ (i.e. each $\widetilde{z}_\ell$ contains ${k-1}$ features), $X^k$ the sequence to be aligned, and $p = (p_1, \ldots, p_L) = \big((n_1,m_1),\ldots,$ $(n_L,m_L)\big)$ an alignment between $\widetilde{Z}$ and $X^k$. %
Intuitively, to add $X^k$ to $\widetilde{Z}$, we use $p$ to stretch $\widetilde{Z}$ and $X^k$ such that corresponding features are aligned and become part of the same element of $Z$. However,
whenever features need to be copied to do the stretching (step sizes $(1,0)$ and $(0,1)$), we rather insert a special \emph{gap} symbol instead of the features themselves. More precisely, let $Z = (z_1,\ldots,z_L)$ denote the updated template, $z_{n}(k)$ denote the $k$-th feature in the $n$-th element of $Z$, and $G$ denote the gap symbol\footnote{Since chroma features contain only non-negative entries, the gap symbol can often be encoded as a pseudo-feature having negative entries.}.
Set $z_1 = (\widetilde{z}_1(1),\ldots,\widetilde{z}_1(k-1),x^k_1)$, then for $l = (2, 3, \ldots, L)$:
\[
z_\ell = 
\begin{cases}
(\widetilde{z}_{n_\ell}(1),\ldots,\widetilde{z}_{n_\ell}(k-1),x^k_{m_\ell}), \;\; p_\ell - p_{\ell-1} = (1,1)\\
(\widetilde{z}_{n_\ell}(1),\ldots,\widetilde{z}_{n_\ell}(k-1),G), \quad\;\; p_\ell - p_{\ell-1} = (1,0)\\
(G,\ldots,G,x^k_{m_\ell}), \quad\quad\quad\quad\quad\;\;\;\; p_\ell - p_{\ell-1} = (0,1)\\
\end{cases}
\]
The gap symbol and its influence will be further discussed in Section~\ref{sec:evaluation}.

The alignment procedure itself is almost identical to standard DTW; only the local cost measure has to be adjusted to take the properties of the template into account. For a template $Z$ comprising $k-1$ feature sequences and a feature sequence X, we define a template-aware cost function $c_T: (\fspace \cup G)^{k-1}\times \fspace \to \R$ as
\[
c_T(z_n,x_m) = \sum_{r=1}^{k-1}
\begin{cases}
c(z_n(r),x_m), & z_n(r) \ne G,\\
{\gappen},  & z_n(r) = G,
\end{cases}
\]
where $\gappen>0$ is a constant referred to as the \emph{gap penalty}.

The influence a single additional recording can have using progressive alignment is illustrated in Fig.~\ref{fig:Motivation}(d). Here, we included a third performance by Poblocka %
in the alignment, which could be considered as being ``between" the two versions shown in Fig.~\ref{fig:Motivation} in terms of articulation style and balance. As we can see, the resulting path (black) follows the ground-truth markings (red dots) quite closely and improves significantly over the pairwise result.

\subsection{Order of Alignments and Iterative Processing}
\label{sec:furtherImprovements}

The alignment of the first two versions in our progressive approach is equivalent to standard pairwise alignment. Errors in this first step influence to some degree all subsequent alignment steps. We discuss now two strategies that can help to increase the reliability of the first few alignments in our progressive approach. First, the order in which the alignments are computed is of importance, and we should start with recordings that are easy to align.
In computational biology, a common approach to identify a reasonable order is referred to as the \emph{guide tree approach} \cite{DurbinEKM99_BiologicalSequenceAnalysis_CUP}.
While there are various ways to implement such an approach, we consider the following procedure.
First, for each pair of recordings, we compute the total cost of an optimal alignment between the pair to identify the pair having the lowest \emph{average cost}, which is defined as the total cost of the alignment divided by its length $L$. We call the feature sequences for the recordings in this pair $X^1$ and $X^2$. 
For the next recording, we identify the one being jointly closest to $X^1$ and $X^2$.
To this end, we sum for each of the remaining recordings the average cost of the alignments between the recording and $X^1$, and the recording and $X^2$. We call the feature sequence of the recording with the lowest sum $X^3$. We continue with this procedure until all recordings are in order. We refer to this strategy as \emph{DTW-cost-based order}.

While this strategy leads to a useful order, its computational costs are significant.
In our experiments, we found an alternative based on a much simpler strategy: We sorted the versions according to their length, starting with the shortest recordings. In the following, we refer to this strategy as \emph{length-based order}. In Section~\ref{sec:evaluation}, we compare both ordering strategies and discuss their behavior.

A second strategy to improve the reliability of the first alignments is referred to as \emph{iterative progressive alignment}. The idea behind this strategy is, after all versions are aligned and included in the template, to remove one version from the template and realign it, starting with the first version that was aligned. This way, errors made early in the progressive alignment can potentially be corrected. We implemented this extension as well and discuss it in Section~\ref{sec:evaluation}.

\subsection{Increasing the Computational Efficiency and Alignment Accuracy}
\label{sec:increasedEffAcc}

Since progressive alignment shares its algorithmic roots with standard DTW, we can incorporate extensions that were successfully used with DTW-based methods.
In particular, the methods described in \cite{MuellerMK06_MsDTW_ISMIR,EwertMG09_HighResAudioSync_ICASSP} employ a variant of DTW referred to as multiscale DTW (FastDTW) to increase the computational efficiency. The general idea is to recursively project an alignment computed at a coarse feature resolution level to a next higher resolution, and to refine the projected alignment on that resolution. This way, the matrix $D$ only has to be evaluated around the projected path. This multiscale approach typically leads to a significant drop in runtime by up to a factor of $30$, see \cite{MuellerMK06_MsDTW_ISMIR}.

Furthermore, the authors in \cite{EwertMG09_HighResAudioSync_ICASSP} introduce a type of features that indicate onset positions separately for each chroma. These chroma-based onset features (DLNCO features) are then combined with normalized chroma features. As shown by the experiments in \cite{EwertMG09_HighResAudioSync_ICASSP}, these combined features can lead to a significant increase in alignment accuracy for pairwise methods. 
In the following, we employ the same features and cost measure as used in \cite{EwertMG09_HighResAudioSync_ICASSP}.

\section{Experiments}\label{sec:evaluation}

To illustrate the performance of our proposed method as well as the influence of certain parameters, we conducted a series of experiments
using recordings taken from the Mazurka Project\footnote{\url{http://www.mazurka.org.uk}},  %
which compiled a database of over $2700$ recorded performances by more than $130$ distinct pianists for $49$ Mazurkas composed by Fr{\'e}d{\'e}ric Chopin.
The recordings are dated between $1902$ and today, and were made under strongly varying recording conditions.
For our experiments, we employ a subset of five Mazurkas and $288$ recordings,
for which manually annotated beat positions are available, see Table~\ref{tab:Recordings}.
Performances with structural differences compared to the majority of recordings (such as additional repetitions of a part of a piece) were excluded from our experiments.

\begin{table}[t]
  \small
	\centering
		\begin{tabular}{llcc}
			\textbf{ID} & \textbf{Piece} & \textbf{No. Rec.} & \textbf{No. Pairs}\\
			M17-4 & Opus 17 No. 4 & 62 & 1891 \\
			M24-2 & Opus 24 No. 2 & 62 & 1891 \\
			M30-2 & Opus 30 No. 2 & 34 & 561 \\
			M63-3 & Opus 63 No. 3 & 81 & 3240 \\
			M68-3 & Opus 68 No. 3 & 49 & 1176 \\              
		\end{tabular}
	\caption{\small Chopin Mazurkas and their identifiers used in
our experiments. The last two columns indicate the number of performances available for the respective piece and the number of evaluated unique pairs.}
	\label{tab:Recordings}
\end{table}

\subsection{Evaluation Measure}
\label{sec:EvaluationMeasure}

To evaluate the accuracy of an alignment between two different versions of a piece, we employ the beat annotations as ground truth. To this end, we use the alignment to locate for each annotated beat position in the one version a corresponding position in the other version. Using the manual beat annotations for the other version, we can then compute the absolute difference between the correct beat position and the one obtained from the alignment. By averaging these differences for all beats, we obtain the \emph{average beat deviation (ABD)} for a given alignment, which we measure in milli-seconds.
For our evaluation, we compute this measure for each Mazurka and each pair of recordings. For example, for M17-4 our setup contains $62$ recordings, which results in $\binom{62}{2}=1891$ unique pairs and corresponding average beat deviation values, see Table~\ref{tab:Recordings}.

\subsection{Pairwise vs Progressive Alignment}
\label{sec:EvalPairwiseVsProgressive}

\begin{figure*}[!t]
 \centering
\includegraphics[width=17cm]{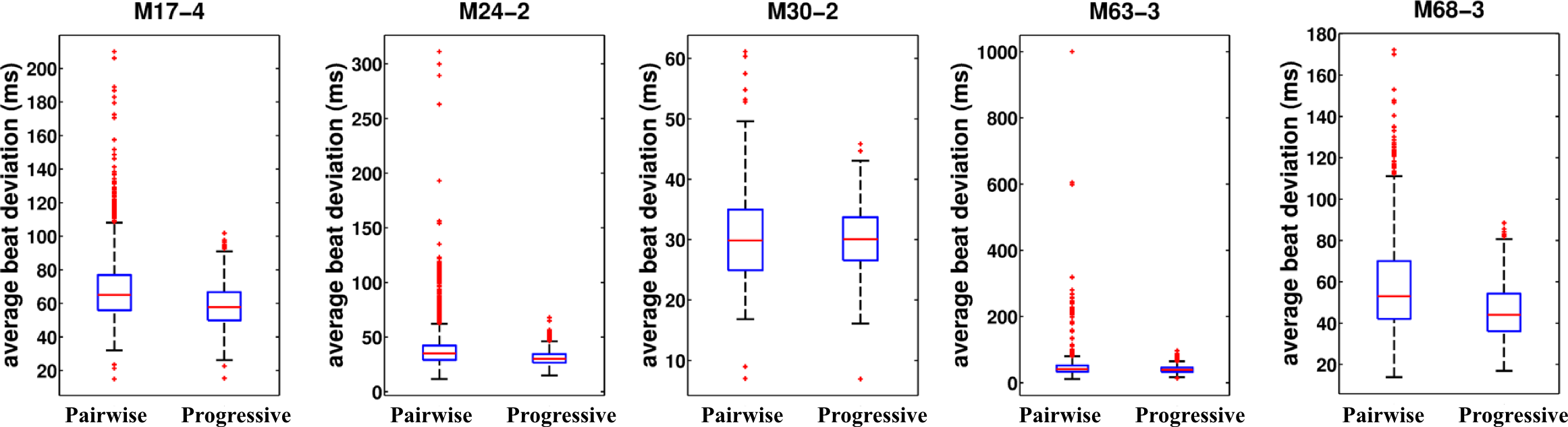}
 \caption{\small Comparison of the baseline pairwise alignment method with our proposed progressive alignment method. The boxplots illustrate the distribution of the average beat deviation values for each Mazurka separately.}
 \label{fig:resultsPairwiseVsProgressive}
\end{figure*}

In a first experiment, we compare the alignment accuracy for pairwise and progressive alignment. Since the pairwise method described in \cite{EwertMG09_HighResAudioSync_ICASSP} employs the same features and cost measure as our proposed progressive method, we use \cite{EwertMG09_HighResAudioSync_ICASSP} as a baseline (other pairwise methods \cite{DixonW05_MATCH_ISMIR} showed a similar behavior). In particular, we use a temporal resolution of 20ms for both chroma and onset-indicator (DLNCO) features. The DTW weights are set to $(w_1 ,w_2 ,w_3) = (2,1.5,1.5)$. As proposed in \cite{EwertMG09_HighResAudioSync_ICASSP}, we use the cosine distance for the chroma features and the Euclidean distance for the DLNCO features. Moreover, for our proposed progressive alignment, we use the length-based alignment order and set the gap penalty parameter to the highest value the cost measure $c$ can assume.
The distribution of the average beat deviation (ABD) values for all pairs is summarized for each of the five Mazurkas separately in the boxplots\footnote{We use standard boxplots: the red bar indicates the median, the blue box gives the $25th$  and $75th$  percentiles ($p_{25}$ and $p_{75}$), the black bars correspond to $p_{25}-1.5(p_{75}-p_{25})$ and $p_{75}+1.5(p_{75}-p_{25})$, and the red crosses are called outliers.}
shown in Fig.~\ref{fig:resultsPairwiseVsProgressive}, as well as in column A and B in Table~\ref{tab:resultsTable}.

Comparing the results for pairwise and progressive alignment, we can see that the mean ABD drops slightly using the progressive approach for most examples. For example, the mean ABD for M17-4 drops from 68ms using pairwise alignment to 59ms using our progressive method (decrease by 13\%). On average, the mean ABD drops by 14\%. More importantly though, the progressive alignment is significantly more stable. In particular, the inter-quartile range is smaller for all five Mazurkas using the progressive alignment (Fig.~\ref{fig:resultsPairwiseVsProgressive}). Further, the number of alignments with a very high ABD is significantly reduced. This can be measured by the standard deviation (std), which for M17-4 using pairwise alignment is 19ms, while progressive alignment leads to an std of 12ms. This difference is even greater for other Mazurkas (M24-2 and M63-3). On average, the std is reduced by more than 50\%. So overall, while our proposed procedure also led to an increase in alignment accuracy on average, the main effect is a gain in robustness against strongly incorrect alignments. 

\begin{table}[!t]
	\centering
	\scalebox{0.95}{
	\small
		\begin{tabular}{lcccccccc}
\textbf{M17-4}&\textbf{[A]}&\textbf{[B]}&\textbf{[C]}&\textbf{[D]}&\textbf{[E]}&\textbf{[F]}&\textbf{[G]}\\
\hline
min&15&15&17&15&15&15&19\\
mean&68&59&68&63&76&80&91\\
max&210&102&118&116&789&129&252\\
std&19&12&13&13&94&13&22\\
\vspace{-0.2cm}&&&&&&&\\
\textbf{M24-2}&&&&&&&\\
\hline
min&12&15&17&12&15&16&11\\
mean&39&31&38&33&31&46&56\\
max&311&68&118&59&68&98&320\\
std&20&6&12&7&6&9&22\\
\vspace{-0.2cm}&&&&&&&\\
\textbf{M30-2}&&&&&&&\\
\hline
min&7&7&7&7&16&6&6\\
mean&30&30&31&29&31&40&43\\
max&61&46&49&53&46&64&80\\
std&8&5&6&6&5&7&9\\
\vspace{-0.2cm}&&&&&&&\\
\textbf{M63-3}&&&&&&&\\
\hline
min&11&13&15&12&13&14&9\\
mean&46&40&46&40&40&53&62\\
max&1000&97&99&99&97&109&1000\\
std&32&11&12&11&11&11&33\\
\vspace{-0.2cm}&&&&&&&\\
\textbf{M68-3}&&&&&&&\\
\hline
min&14&17&21&15&17&21&12\\
mean&58&46&57&53&46&71&86\\
max&172&89&144&105&89&179&335\\
std&23&13&18&15&13&21&34\\
		\end{tabular}
		}
\caption{\small Statistics over the average beat deviation (ABD) values for the five Mazurkas and for 7 different alignment approaches (see text).
\textbf{[A]}: Pairwise alignment.
\textbf{[B]}: Proposed progressive alignment.
\textbf{[C]}: Proposed without gap symbols.
\textbf{[D]}: Proposed using DTW-cost-based alignment order.
\textbf{[E]}: Proposed using iterative alignment.
\textbf{[F]}: Proposed without DLNCO features.
\textbf{[G]}: Pairwise without DLNCO features.
All values in milli-seconds.
}
\label{tab:resultsTable}
\end{table}

\subsection{Gap Penalties}
\label{sec:EvalGap}

In the next experiment, we investigate the influence of the gap penalty parameter by testing a slightly modified version of our proposed method. To this end, we modify the way the template is creating by setting $z_\ell = (\widetilde{z}_{n_\ell}(1),$ $\ldots,\widetilde{z}_{n_\ell}(k-1),x^k_{m_\ell})$ for $\ell \in$ \setmacro{1}{L}, i.e. we do not insert gap symbols but copy features as necessary to create the new template (comparing to Section~\ref{sec:jointAlignment}). The results using this modification are shown in column C in Table~\ref{tab:resultsTable}. Comparing these values to our proposed method (column B) and the reference pairwise method (column A), we see that this gap-less version typically improves over pairwise alignment in terms of maximum ABD values and the standard deviation, just as the proposed method. For example, for M17-4, the max ABD in column A is 210ms, while the max ABD in column C is 118ms. However, we do not observe a decrease in the mean ABD compared to pairwise alignment. For example, for M17-4, while using gaps the mean ABD drops from 68ms (column A) to 59ms (column B), it stays on a similar level in column C (68ms). The reason could be that by copying the features to create the template, some temporal precision is lost and this results in a minor loss of alignment accuracy. 

\subsection{Alignment Order}
\label{sec:EvalAlignmentOrder}

Next, we investigate the influence of the order in which we compute the progressive alignment, comparing the length-based and the DTW-cost-based strategy (see Section~\ref{sec:furtherImprovements}). The results are given in columns B and D of Table~\ref{tab:resultsTable}, respectively. As we can see, there are no significant differences between both strategies.
 For example, for M17-4, the mean ABD using the length-based strategy is 59ms (column B), while using the DTW-cost-based strategy the ABD slightly increases to 63ms. The other statistical values show a similar behavior.
Since these results do not disclose any obvious advantages for the DTW-cost-based strategy, we therefore propose to simply use the length-based strategy. 
Interestingly, using the length-based strategy but starting with the longest recordings led to worse results.

Since (local) tempo differences can usually be handled quite well using DTW, it is not obvious why sorting by length yields a useful order. However, the fact that it does 
could indicate that there might be a correlation between the chosen tempo and other expressive parameters, such as articulation or balance, as strong differences in these parameters typically lead to difficulties for the alignment. Furthermore, the fact that according to our evaluation the shorter recordings were easier to align, could indicate that a high tempo could limit the range of possible realizations of expressive parameters in a performance. However, further studies would be necessary to confirm such theories.

\subsection{Iterative Alignment}
\label{sec:EvalIterative}

In a further experiment, we investigate whether iterative processing could further improve the alignment accuracy, compare Section~\ref{sec:furtherImprovements}.
To this end, we use two iterations: the first iteration corresponds to progressive alignment, and in the second iteration, each version is removed from the template once and is then realigned. 
The results for this extension are given in column E of Table~\ref{tab:resultsTable}. Overall, the iterative variant led to a slight decrease in ABD in almost all examples, which is not even visible in Table~\ref{tab:resultsTable} as we rounded all values. On the contrary, we observed a significant increase in ABD for M17-4 using the iterative variant. Here, the realignment led to a misalignment of several shorter recordings.
Therefore, the results do not indicate any significant advantages of using iterative alignment. 

\subsection{Influence of Onset-Indicator Features}
\label{sec:EvalDLNCO}

In a final experiment, we investigate the influence of the chroma-based onset-indicator (DLNCO) features\cite{EwertMG09_HighResAudioSync_ICASSP} on the alignment accuracy when using progressive alignment. To this end, we disabled the DLNCO features in our proposed method, and computed the alignment only based on the normalized chroma features. The results of this experiment are given in column F in Table~\ref{tab:resultsTable}. As a further reference, we disabled the DLNCO features in our baseline pairwise method as well (column G).

As we can see, the minimum over the ABD values remains unaffected for most of the Mazurkas, which means that easy to align pairs can be aligned with chroma features alone just as well. For example, for M17-4, the minimum value in column F is identical to the one in column B. However, we see a significant increase in ABD in all other statistical values. For example, the mean ABD for M17-4 for our proposed method including DLNCO features is 59ms (column B), while disabling the DLNCO leads to a mean ABD of 80ms (column F). Similar observations can be made comparing the pairwise results.
Overall, the results seem to indicate that including onset-indicator features indeed leads to a significant increase in alignment accuracy also for progressive alignments.
\section{Conclusion}\label{sec:conclusion}

In this paper, we introduced a method for aligning multiple versions
of a piece of music in a joint way. The availability of multiple
versions to compare against during the alignment, stabilized the comparison
for hard-to-align recordings and led to an overall increase in alignment
accuracy and, in particular, in alignment robustness. Our experiments using
real-world recordings from the Mazurka Project demonstrated that our proposed method
can indeed be used to raise the alignment accuracy compared to previous methods that
are limited to pairwise alignments. For the future, we plan to further investigate
the behaviour of our procedure. In particular, we plan to analyze how other
ordering strategies influence the alignment accuracy. We will also further explore
different strategies to implement a cost for the gap symbol and to make it more 
adaptive. 

{\small \noindent \textbf{Acknowledgements:} This work was partly funded by the China Scholarship Council (CSC), EPSRC Grant EP/J010375/1, and the Queen Mary Postgraduate Research Fund (PGRF).}

\bibliography{referencesNew,referencesMusic}

\end{document}